\documentclass{iau}
\usepackage{graphicx}

\newcommand{\AD}{\mbox{ATLAS$^{\rm 3D}$}}

\title[Deep imaging of massive galaxies] 
{Probing the mass assembly of massive nearby galaxies with deep imaging}

\author[Duc et al.]{P.--A.\ Duc,$^1$  J.-C. Cuillandre,$^2$ K.\ Alatalo,$^3$ L.\ Blitz,$^3$ M.\ Bois,$^4$ F.\ Bournaud,$^1$ M.\ Bureau,$^5$  M.\ Cappellari,$^5$  P. C\^ot\'e,$^{6}$ R.\ L.\ Davies,$^5$ T.\ A.\ Davis,$^7$ P.\ T.\ de~Zeeuw,$^{7,8}$ E.\ Emsellem,$^{7,4}$ L.\ Ferrarese,$^6$   E.\ Ferriere,$^1$ S. Gwyn,$^{6}$  S.\ Khochfar,$^{9}$ D.\ Krajnovic,$^7$ H.\ Kuntschner,$^7$ P.-Y.\ Lablanche,$^4$ R.\ M.\ McDermid,$^{10}$ L.\ Michel-Dansac,$^4$ R.\ Morganti,$^{11}$ T. Naab,$^{12}$ T.\ Oosterloo,$^{11}$ M.\ Sarzi,$^{13}$ N.\ Scott,$^5$ P.\ Serra,$^{11}$  A.\ Weijmans$^{14}$ and L.\ M.\ Young$^{15}$}

\affiliation{
$^1$ AIM Paris-Saclay, France;
$^2$ CFHT, USA;
$^3$University of California, Berkeley, USA;
$^4$Observatoire de Lyon, France;
$^5$University of Oxford, UK;
$^6$Herzberg Institute of Astrophysics, Victoria, Canada;
$^7$ESO, Garching, Germany;
$^8$Leiden University, The Netherlands;
$^{9}$MPE, Garching, Germany;
$^{10}$Gemini Observatory, Hilo, USA;
$^{11}$ASTRON, Dwingeloo, The Netherlands;
$^{12}$ MPI, Garching, Germany;
$^{13}$University of Hertfordshire, Hatfield, UK;
$^{14}$Dunlap Institute for Astronomy \& Astrophysics,  University of Toronto, Canada;
$^{15}$New Mexico Tech, Socorro, USA
}

\pubyear{2013}
\volume{295}  
\pagerange{xx--xx}
\jname{The intriguing life of massive galaxies}
\editors{D. Thomas, A. Pasquali \& I. Ferreras, eds.}
\begin{document}

\maketitle

\begin{abstract}
According to a popular scenario supported by numerical models, the   mass assembly and growth of massive galaxies, in particular the Early-Type Galaxies (ETGs), is, below a redshift of 1,  mainly due to the accretion of multiple gas--poor satellites. In order to get observational evidence of the role played by minor dry mergers,  we are obtaining extremely deep optical images of a complete volume limited sample of  nearby ETGs. 
 These observations, done with the CFHT as  part of the \AD, NGVS and MATLAS  projects,   reach a stunning 28.5 -- 29 mag.arcsec$^{-2}$ surface brightness limit in the g'  band. They allow us to detect the relics of past collisions such as faint stellar tidal tails as well as the very extended  stellar halos which keep the memory of the last episodes of galactic accretion.   
  Images and preliminary results from this on-going survey are presented, in particular a possible correlation between the fine structure index (which parametrizes the amount of tidal perturbation) of the ETGs, their stellar mass, effective radius and gas content.

\end{abstract}

\firstsection 
\section{Introduction}

Early-Type Galaxies
  play a key role in modern cosmology: according to the  standard hierarchical cosmological model,  galaxies build up from successive mergers, associated by a series of morphological transformations. The massive ETGs we see today are the end-product of this process.
{\it At high redshifts}, few ETGs are observed, but surprising  they have not all disappeared, raising questions on how they  formed so quickly in the traditional merging scenario.  They appear to be also very compact (e.g. Buitrago et al., 2008).
{\it At low redshift},  the  early-type galaxies are observed to be larger but also  to be much more complex and lively that once believed. As  shown  in the presentations of the \AD\ results in this volume,   a large fraction of them  contain in particular gas,  some with non-regular kinematics, and thus may still be involved in  active transformation processes. 

Simulations indicate that while the global  light profile  of  ETGs  can be rather easily obtained with various processes, getting their total mass  and large radius is much more challenging and may require multiple collisions in the recent past.  Several  recent papers have  highlighted the role of  minor mergers in the growth of galaxies (e.g., Johansson et al., 2012; Newman et al., 2012).
These late mergers impact the properties of the stellar populations mostly at large galactocentric radii. In particular minor mergers bring  low metallicity stars from the infalling dwarf satellites,  and create radial color gradients. On the other hand, a major merger induces large radial mixing, leading to a washing up of metallicity/age ad thus color gradients.

Furthermore, the mass assembly of galaxies  leaves various imprints in their surroundings, such as shells, streams and  tidal tails. The  frequency, shape and properties  of these fine structures depend on the mechanism driving the mass assembly (see the review by Duc \& Renaud, 2013).  Any analysis of the fine structures around galaxies, however, should take into account (1) that the resulting stellar debris   have a very low surface brightness, (2) that such debris   fade with age, or may be dispersed by the local environment, and (3)  that galaxies may have formed by multiple processes.
 
   Deep imaging surveys coupled with numerical simulations done in cosmological context can now address these issues. 
Several studies have quantified  the importance of fine structures around massive galaxies and their variation with environment and redshift  (Tal et al., 2009; Bridge et al., 2010; Adams et al., 2012) but were restricted to the census of luminous tidal features.  
So far, the  surface brightness limit required to apply  a genuine galactic archeology technique has been reached for only local galaxies for which stellar counts may be done. As however shown, among others,  by  Martinez-Delgado et al. (2010) and Duc et al. (2011),  deep optical images obtained under specific  conditions can also reveal the diffuse light associated with very low-surface brightness structures.

\section{Observations and data reduction}

Our targets are the 260 nearby ETGs located  at distances below 42 Mpc  from the \AD\ survey (Cappellari et al., 2011).  
Very deep optical images in the g',r' and i' bands  are currently being obtained with the large field of view camera MegaCam installed on the CFHT. The observations are carried out  as part of several projects: \AD,  the Next Generation Virgo Cluster Survey (NGVS, covering the full Virgo Cluster area;  Ferrarese et al., 2012), and the  recently accepted CFHT MATLAS Large Programme. The typical integration time is about one hour per band. The limiting sensitivity is outstanding with respect to previous generations of   optical images: about 28.5 -- 29 mag.arcsec$^{-2}$ in the g'--band. This could be achieved using  dedicated observing strategies and data reduction softwares.  
On traditional images obtained by MegaCam, the presence of scattered light patterns masks extended features below surface brightness of 27 mag.arcsec$^{-2}$. 
Investigations motivated by the NGVS have shown that this problem can be overcome carrying out  a sequence of observations with large offsets between the images, as it is usually done with infrared observations. A sky is then computed and subtracted from the individual images, before they are stacked. 

The majority of the fine structures disclosed by the survey are very extended and directly show up on surface brightness  or color maps. To disclose those located  more towards  the inner regions, several  methods were used:   unsharp masking, best at revealing  sharp-edged structures such as shells or narrow filaments; subtraction of the ETGs modeled by an ellipse fitting or by a multi-component GALFIT model, which helps to detect extended asymmetric features. As a first step to quantify the amount of tidal perturbations, a fine structure index was determined, counting by eye the number of structures, and weighting  them according to their nature.

\section{Preliminary results}
\begin{figure}[b]
\begin{center}
 \includegraphics[width=\textwidth]{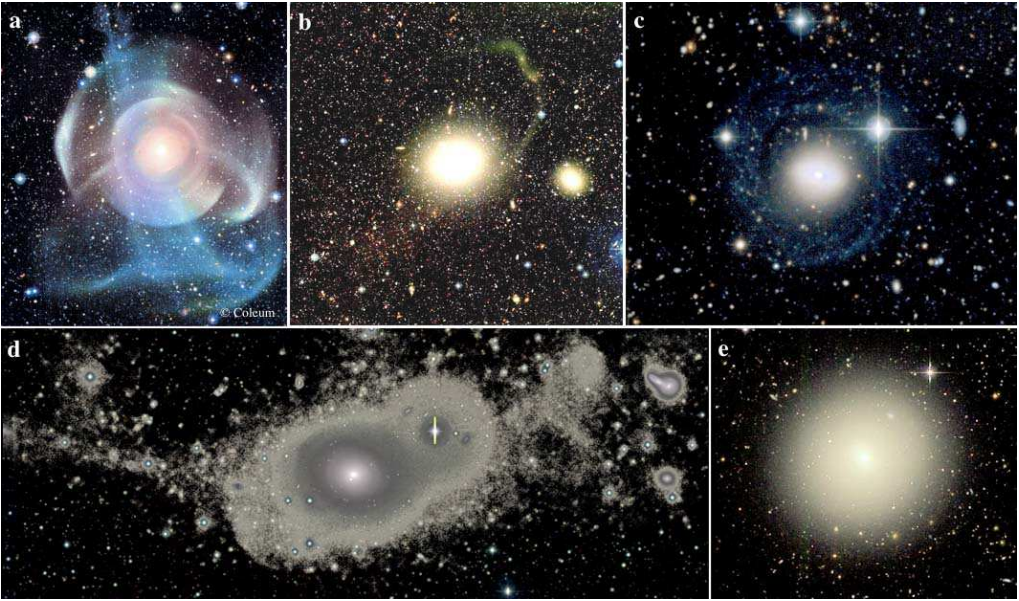} 
 \caption{Deep optical images of a sub-sample of nearby Early-Type Galaxies obtained with MegaCam on the CFHT as part of the \AD\ and Next Generation Virgo Cluster Survey. The figure illustrates the variety of low surface brightness structures that show up around these galaxies:
 long tidal tails and shells, telling us about past major mergers (a,d); narrow stellar filaments associated with disrupted dwarf satellites, revealing future minor mergers (b); regular low surface brightness star--forming disks (c);  extended featureless  stellar halos (e).
  }
\label{fig1}
\end{center}
\end{figure}

Figure~\ref{fig1} illustrates the variety of low surface brightness structures detected around the ETGs: 200 kpc long tidal tails revealing past 2-3 Gyr old  major merger (for instance   around NGC~5557; Duc et al., 2011), narrow filaments around a disrupted dwarf, with their typical S-shape and  wrapping  around the host, diffuse halos, some remaining regular even up to large radii.

At time of writing, more than half of the \AD\  ETGs benefit from deep MegaCam images. This initial sub-sample is however somehow biased  towards the most massive ETGs, those that are slow rotating and/or  gas--rich. In such conditions, providing the percentage of disturbed galaxies  is  premature.  Nevertheless some initial trends were found. 

\begin{figure}[b]
\begin{center}
\includegraphics[height=0.9\textwidth,width=7.2cm,angle=-90]{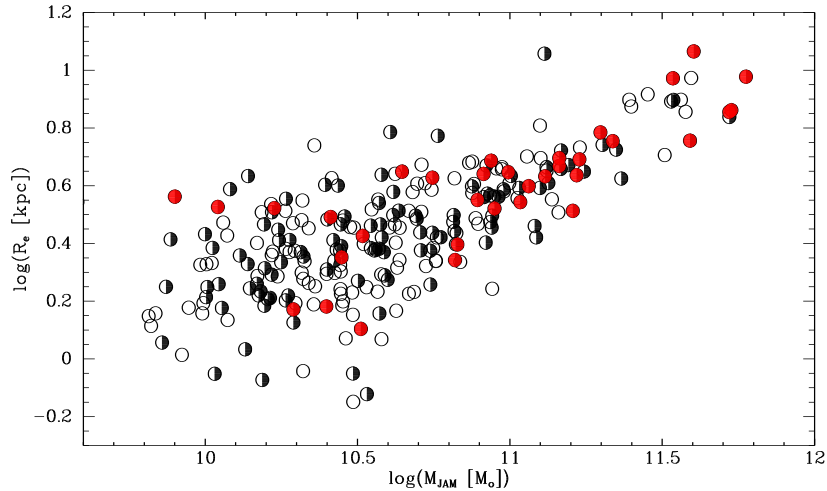} 
 \caption{Effective radius, $R^{\rm{max}}_e$, vs M$_{\rm{JAM}}$, a proxy of the stellar mass, for the full sample of  \AD\ ETGs (open circles, Cappellari et al., 2012). The galaxies shown with the half filled circles have already deep MegaCam images available from either the \AD\ or NGVS surveys.  The images for the  other galaxies will be obtained as part of the MATLAS CFHT Large Programme. Those that have a high fine structure index, i.e. galaxies that are either strongly tidally perturbed, or have multiple stellar streams in their vicinity,  are shown with the red filled circles.}
\label{fig2}
\end{center}
\end{figure}

First of all, statistically, the galaxies that contain atomic hydrogen, in particular in their outskirts, have a higher fine structure index. This is a strong indication that the HI clouds are associated with collisional debris. A rather large fraction of  \AD\  ETGs contain detectable molecular gas, as traced by CO. Those for which the CO  is kinematically misaligned with respect to the stellar component and for which  an external origin of the molecular gas had been proposed (Davis et al, 2011), exhibit collisional debris, just as expected.
Finally  one of the most promising results is the trend with mass and size, two fundamental parameters in the scaling laws of galaxies. As shown on Figure~\ref{fig2}, the more massive and the more extended\footnote{As mentioned a number of times during the symposium, the measure of the effective radius, the radius containing half of the stellar light, is tricky at high redshift. Our study indicates that this is even the case at low redshift.
For the  \AD\ ETGs showing on MegaCam images a very extended low surface brightness halo, the value had to be revised, with differences with earlier published values  reaching factors up to 2.} the galaxy is, the higher its fine structure index. Conversely, many, usually fast rotating low--mass,  ETGs do not show any sign of external perturbation. 
This tells that the most massive galaxies in the local Universe have had a rich recent mass accretion history. At which level  it  accounts for their mass/size  growth remains to be determined. Since the optical images only reveal structures brighter than 29  mag.arcsec$^{-2}$, simulations will be key to extrapolate from the observations the quantity of accreted material since $z=1$. Another crucial parameter will be the survival time of tidal features, which likely depends on the environment. The study of the  fine structure index as a function of the local density  -- our volume limited sample covers a large range of environments: field, groups and the Virgo Cluster -- will give interesting constrains on that matter.

\end{document}